\documentclass[journal]{IEEEtran}

\usepackage{graphicx}
\usepackage{epsfig}
\usepackage{amssymb}
\usepackage{lineno}
\usepackage{amssymb}
\usepackage[utf8]{inputenc}
\usepackage{array}
\usepackage[above]{placeins}
\usepackage{amsmath}
\usepackage{url}
\usepackage{mathtools}
\usepackage{mathcomp}
\usepackage{tabularx}
\usepackage{multirow}
\usepackage{chngcntr}
\usepackage{algorithm}
\usepackage[noend]{algpseudocode}
\usepackage{xtab,afterpage}
\usepackage[colorlinks]{hyperref}
\usepackage[justification=centering]{caption}
\usepackage{subcaption}
\usepackage{adjustbox}
\usepackage{xspace}
\usepackage{acronym}
\usepackage{longtable}
\usepackage{chngcntr}
\usepackage{enumitem}
\usepackage{braket}

\newcolumntype{P}[1]{>{\hspace{0pt}}p{#1}}
\newcolumntype{C}[1]{>{\centering\arraybackslash}p{#1}}
\counterwithout{paragraph}{subsubsection} 
\counterwithin{paragraph}{subsection} 
\newcounter{magicrownumbers}

\DeclarePairedDelimiter\abs{\lvert}{\rvert}%

\hypersetup{
colorlinks = true,
allcolors = magenta,
hyperfigures = true,
pdfauthor = {Mrinmoy Bhattacharjee, S.R.M. Prasanna, Prithwijit Guha},
pdfcreator = {Mrinmoy Bhattacharjee, Dept. of EEE, IIT Guwahati, India},
pdftitle = {Time-Frequency Audio Features for Speech/Music Classification},
pdfdisplaydoctitle = true,
bookmarksnumbered = true,
}

\begin{document}
\title{Time-Frequency Audio Features for Speech-Music Classification}

\author{Mrinmoy~Bhattacharjee,~\IEEEmembership{Student MIEEE}, 
S.R.M.~Prasanna,~\IEEEmembership{SMIEEE},
Prithwijit~Guha,~\IEEEmembership{MIEEE}
\thanks{}
\thanks{Mrinmoy Bhattacharjee, S.R.M. Prasanna and P. Guha are with the Dept. of Electronics and Electrical Engineering, Indian Institute of Technology Guwahati, Guwahati-781039, India}
\thanks{S.R.M. Prasanna is also with the Dept. of Electrical Engineering, Indian Institute of Technology Dharwad, Dharwad-580011, India}
\thanks{email\{mrinmoy.bhattacharjee,prasanna,pguha\}@iitg.ac.in}
}

\markboth{}%
{Shell \MakeLowercase{\textit{}}: }

\maketitle

\begin{abstract}
Distinct striation patterns are observed in the spectrograms of speech and music. This motivated us to propose three novel time-frequency features for speech-music classification. These features are extracted in two stages. First, a preset number of prominent spectral peak locations are identified from the spectra of each frame. These important peak locations obtained from each frame are used to form Spectral peak sequences (SPS) for an audio interval. In second stage, these SPS are treated as time series data of frequency locations. The proposed features are extracted as periodicity, average frequency and statistical attributes of these spectral peak sequences. Speech-music categorization is performed by learning binary classifiers on these features. We have experimented with Gaussian mixture models, support vector machine and random forest classifiers. Our proposal is validated on four datasets and benchmarked against three baseline approaches. Experimental results establish the validity of our proposal.
\end{abstract}


\IEEEpeerreviewmaketitle

\vspace*{-\baselineskip}
\section{Introduction}
\label{sec:intro}

\IEEEPARstart{C}{ontent} based audio indexing and retrieval applications often involve an important preprocessing step of segmenting and classifying audio signals into distinct categories. Apart from general environmental sounds, speech and music are two important audio categories. Preprocessing steps necessarily require classification algorithms that ensure homogeneity of the category in audio segments. This work focuses on proposing features for better discrimination of speech and music for such audio segmentation applications.

\begin{figure}[t]
\centerline{
\includegraphics[width=0.5\columnwidth]{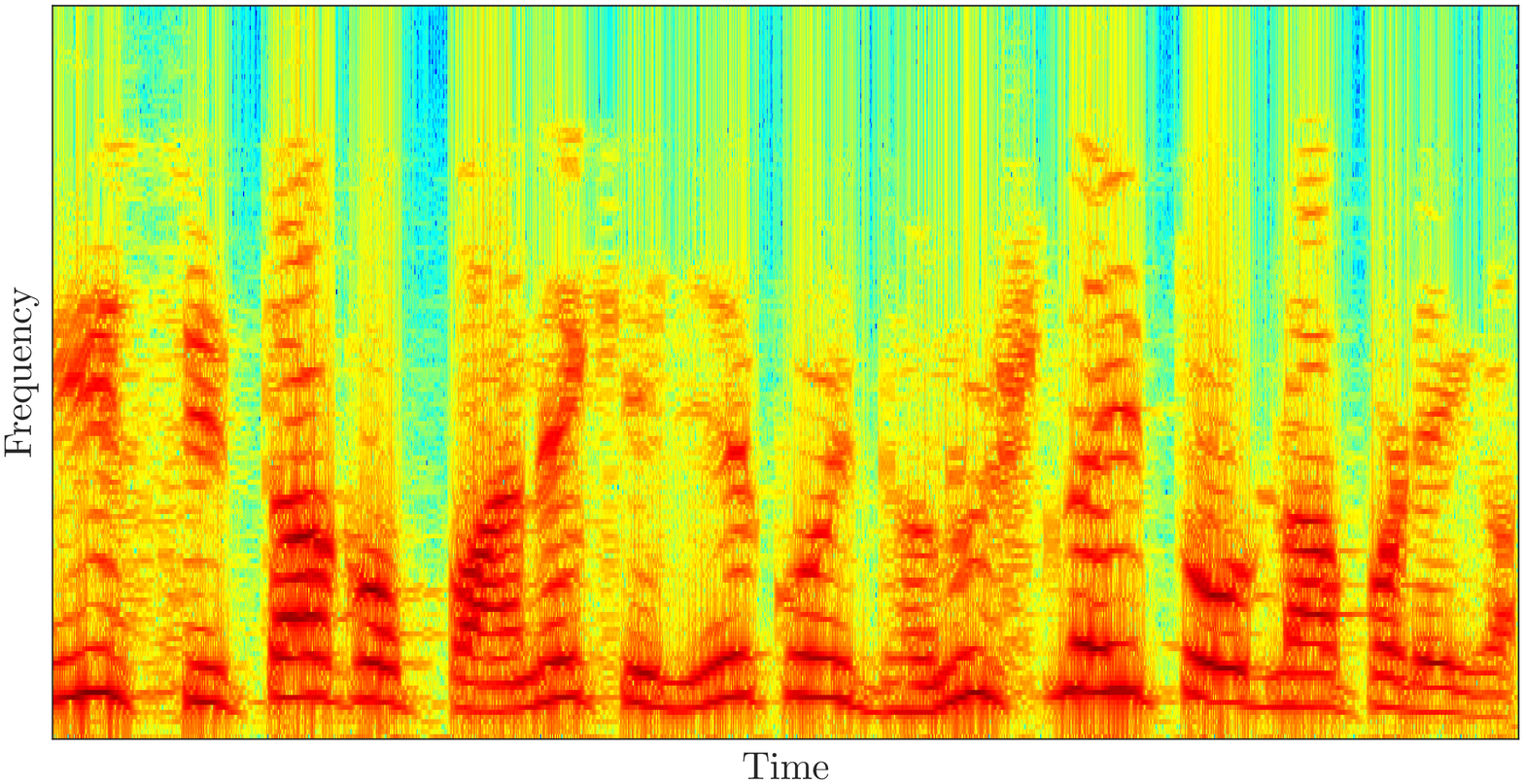}
\includegraphics[width=0.5\columnwidth]{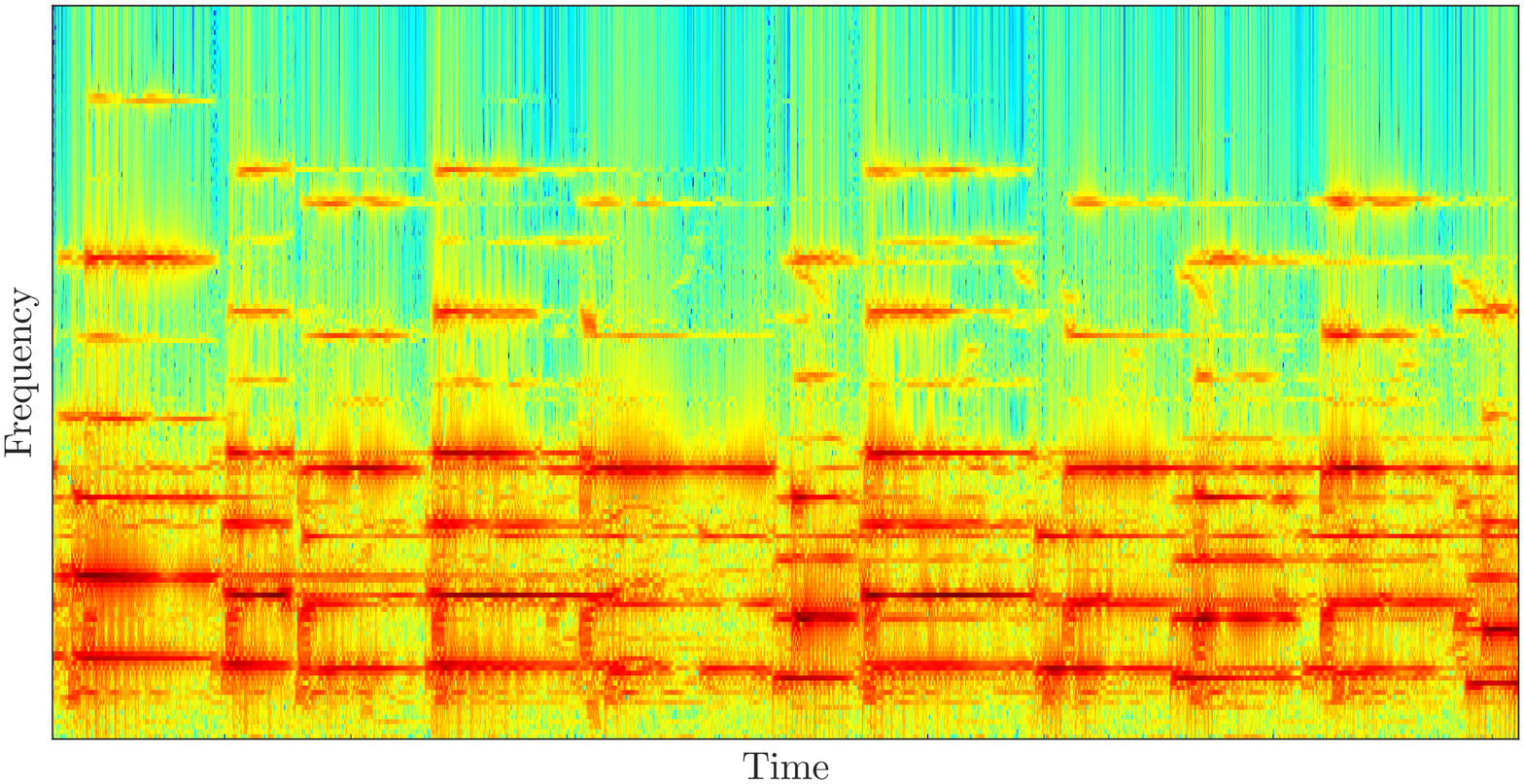}
\vspace{-0.3\baselineskip}
}
\centerline{
(a) \hspace{0.5\columnwidth} (b)
\vspace{-0.3\baselineskip}
}
\setlength{\belowcaptionskip}{-\baselineskip}
\caption{Spectrograms of (a) Speech and (b) Music. Note the distinct striation patterns of speech and music. This observation motivated our proposal of time-frequency audio features for speech-music discrimination.}
\label{fig:spgSM}
\end{figure}

Researchers have observed several differences in speech and music signals. For example, pitch in speech usually exists over a span of $3$ octaves only, whereas music consists of fundamental tones spanning up to $6$ octaves \cite{Saunders_ICASSP1996}. Also, specific frequency tones play an important part in the production of music. Hence, unlike speech, music is expected to have strict structures in the frequency domain \cite{Sell_ICASSP2014}. Furthermore, short silences usually punctuate speech sound units \cite{Panagiotakis_TM2005}, while music is generally continuous and without breaks (Figure~\ref{fig:spgSM}). Literature in the \textit{classification of speech and music} (CSM, henceforth) includes many studies that exploit such (and other) differences between them \cite{Velayatipour_IJCSNS2014,Lavner_EURASIP2009}. We briefly review a few closely related works next.

Table~\ref{table:LiteratureReview} lists the most widely used feature sets of CSM literature. We have categorized these features into two groups viz. \emph{Spectral Features} and \emph{Temporal Features}. Most widely used features from the spectral group are Zero-Crossing Rate (ZCR, henceforth) \cite{Sell_ICASSP2014}, Spectral Centroid, Spectral Roll-off and Spectral Flux \cite{Mezghani_AICCSA2016}. Energy \cite{Neammalai_APSIPA2014}, Entropy \cite{Srinivas_ICDSP2014} and Root Mean Square (RMS) \cite{Sell_ICASSP2014} values are the most popular ones from the temporal group. Apart from these, few works have used spectrograms as features and processed them as images. For example, the approach proposed by Mesgarani et al. \cite{Mesgarani_TASLP2006} is inspired by auditory cortical processing and uses Gabor-like spectro-temporal response fields for feature extraction from spectrogram. On the other hand, Neammalai et al. \cite{Neammalai_APSIPA2014} performed thresholding and smoothing on standard spectrograms to form binary images and used them as features for classification. Existing works on speech-music classification have mostly employed Gaussian Mixture Models (GMM) \cite{Sell_ICASSP2014, Khonglah_DSP2016, Zhang_ISSPIT2016}, Artificial Neural Networks (ANN) \cite{Srinivas_ICDSP2014}, k-Nearest Neighbors (kNN) \cite{Barbedo_JAES2006, Alexandre_EUROCON2005, Burred_JAES2004} and Support Vector Machines (SVM) \cite{Zhang_ISSPIT2016, Mezghani_AICCSA2016, Neammalai_APSIPA2014} as classifiers. Recent works have also used deep learning techniques for this task \cite{Kruspe_Informatik2017,Pikrakis_EUSIPCO2014}. Most existing works have attempted to characterize speech or music using pure temporal and/or spectral features. We believe that time-frequency feature based representations are necessary for better speech-music classification. Our motivation for this proposal is described next.

Figure~\ref{fig:spgSM} shows the spectrograms of speech and music. In case of speech, pitch and harmonics slowly change from one frame to another \cite{Yi_JASA2002}. This leads to the formation of smooth arc-like patterns in its spectrogram. On the other hand, pitch and harmonics in music remain stationary for some finite duration before performing sharp transitions \cite{Jeremy_SIAM2002}. As such, music spectrograms contain patterns in the form of many horizontal line segments. These can be attributed to the following reasons.

\begin{description}[style=unboxed,leftmargin=0cm]
\item[Inertia of speech production system]\label{Inertia} -- Speech production system possesses inertia \cite{Murty_TASLP2008, Zhang_JASA2016}. It requires a finite amount of time to change from one sound unit to another, leading to the formation of slowly changing striation patterns in speech spectrogram. Whereas, individual notes of music have a specific onset instant, marked by a relatively large burst of energy that make its striation patterns discontinuous \cite{Bello_TSAP2005}.

\item[Slowly decaying harmonics in music]\label{Decay} -- Music tones decay slowly. Comparatively, speech production system is a damped system where sound decays quite fast \cite{Meyer_MAAEMAMIM2009,Oller_2008}.

\item[Range of sounds produced]\label{Sounds} -- A musical instrument produces only a fixed number of tones and their overtones. On the other hand, speech production system generates a large number of intermediate frequencies while transitioning from one sound unit to another \cite{Khonglah_TENCON2016, Sell_ICASSP2014}.

\end{description}

The tempo-spectral properties of speech and music are quite distinct. Hence, features capturing joint variations in temporal and spectral domains should be harnessed for efficient classification of speech and music. Existing works in this area have used combinations of temporal and spectral audio features \cite{Sell_ICASSP2014, Mezghani_AICCSA2016, Srinivas_ICDSP2014, Khonglah_DSP2016, Lim_IETSP2012} for achieving better performance. 

We propose three new audio features capable of capturing the joint tempo-spectral characteristics of an audio segment. Peaks in the spectra of audio frames appear as striation patterns in spectrograms. Prominent spectral peaks having relatively higher amplitudes correspond to the brightest patterns in spectrograms. We believe that the frequency locations of such prominent peaks carry class specific information. Accordingly, We compute the features in a two-stage approach. First, these prominent spectral peaks are identified in all frames of an audio interval. Second, locations of detected peaks across frames are treated as temporal sequences, defined as spectral peak sequences (SPS). The proposed features are derived as zero crossing rate, periodicity and second order statistics of each SPS. The speech-music classification is performed by training classifiers on these features. The proposed scheme for feature extraction is described in further detail in Section~\ref{sec:ProposedFeatures}.

We have benchmarked our proposal on four audio datasets and against three baseline approaches \cite{Khonglah_DSP2016, Sell_ICASSP2014, Mezghani_AICCSA2016}. The results of our experiments are reported in Section~\ref{sec:Experiments}. Finally, we conclude in Section~\ref{sec:Conclusion} and sketch the possible future extensions of the present proposal.

\begin{table}[t]
\caption{Most widely used audio features in speech vs music classification literature}
\label{table:LiteratureReview}
\renewcommand{\arraystretch}{1.3}
\resizebox{\columnwidth}{!}{%
\centering
\begin{tabular}{
C{0.15\columnwidth}
C{0.65\columnwidth}
C{0.2\columnwidth}}

\hline
\textbf{Group} & \textbf{Features} & \textbf{Papers} \\

\hline
\textbf{Spectral Features} & ZCR, Spectral Centroid, Spectral Flux, Spectral Rolloff, MFCC, Chroma, Log Mel spectrum energy, Harmonic ratio, Modulation spectrum energy, Pitch & \cite{Khonglah_DSP2016, Mezghani_AICCSA2016, Khonglah_INDICON2015, Srinivas_ICDSP2014, Sell_ICASSP2014, Gallardo_SPL2010, Lavner_EURASIP2009, Pikrakis_TM2008} \\

\hline
\textbf{Temporal Features} & Energy, Entropy, RMS, Peak-to-Sidelobe ratio (PSR) from the Hilbert Envelope of the LP Residual, Normalized Autocorrelation Peak Strength (NAPS) of Zero frequency filtered signal & \cite{Khonglah_DSP2016, Mezghani_AICCSA2016, Srinivas_ICDSP2014, Neammalai_APSIPA2014, Sell_ICASSP2014, Lim_IETSP2012, Lavner_EURASIP2009, Song_SPL2008, Pikrakis_TM2008} \\

\hline
\end{tabular}
}
\vspace{-\baselineskip}
\end{table}

\vspace*{-0.5\baselineskip}

\section{Proposed work}
\label{sec:ProposedFeatures}

\begin{figure*}[tb!]
\centerline{
\includegraphics[width=0.2\textwidth,keepaspectratio]{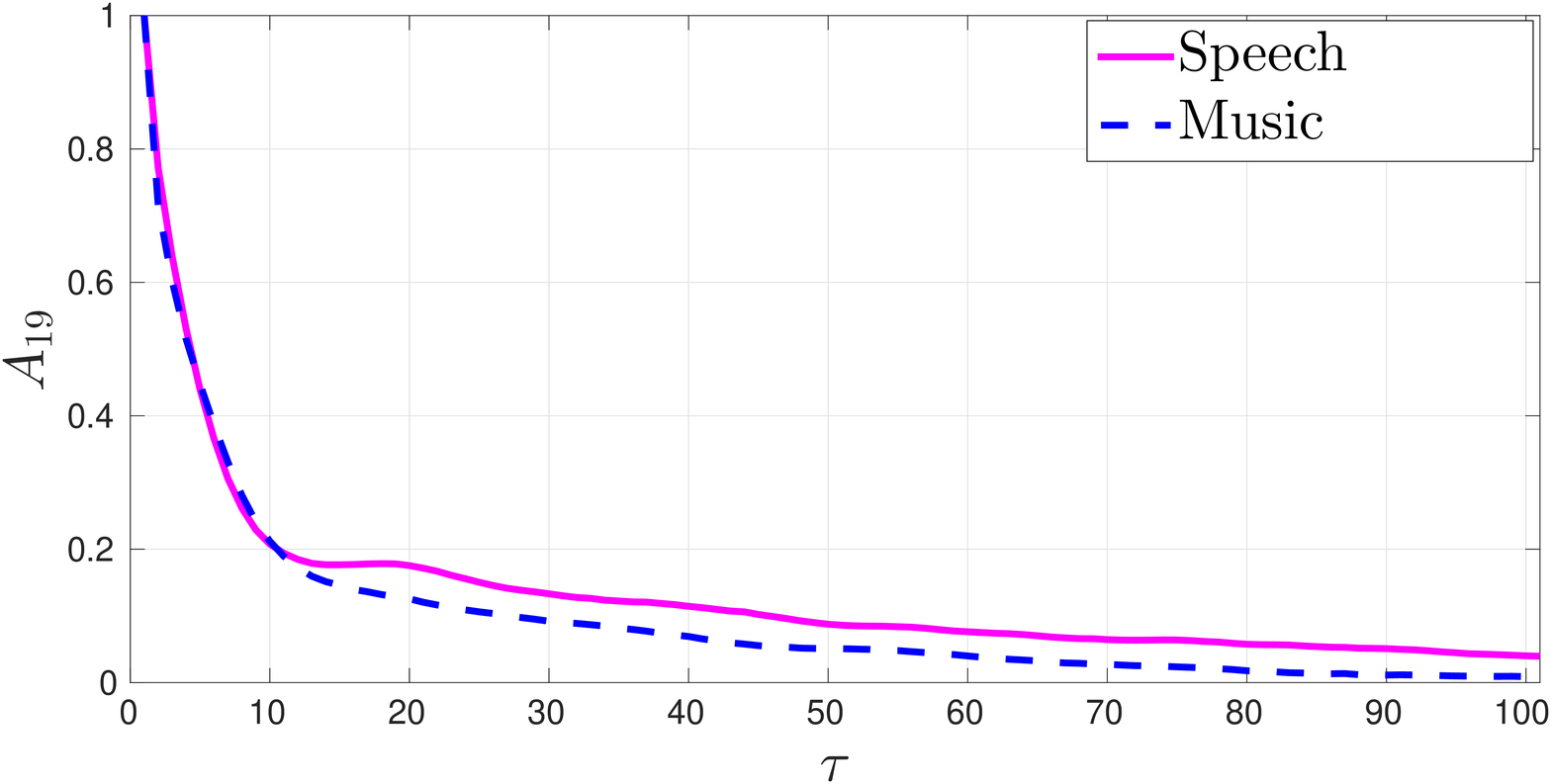}
\includegraphics[width=0.2\textwidth,keepaspectratio]{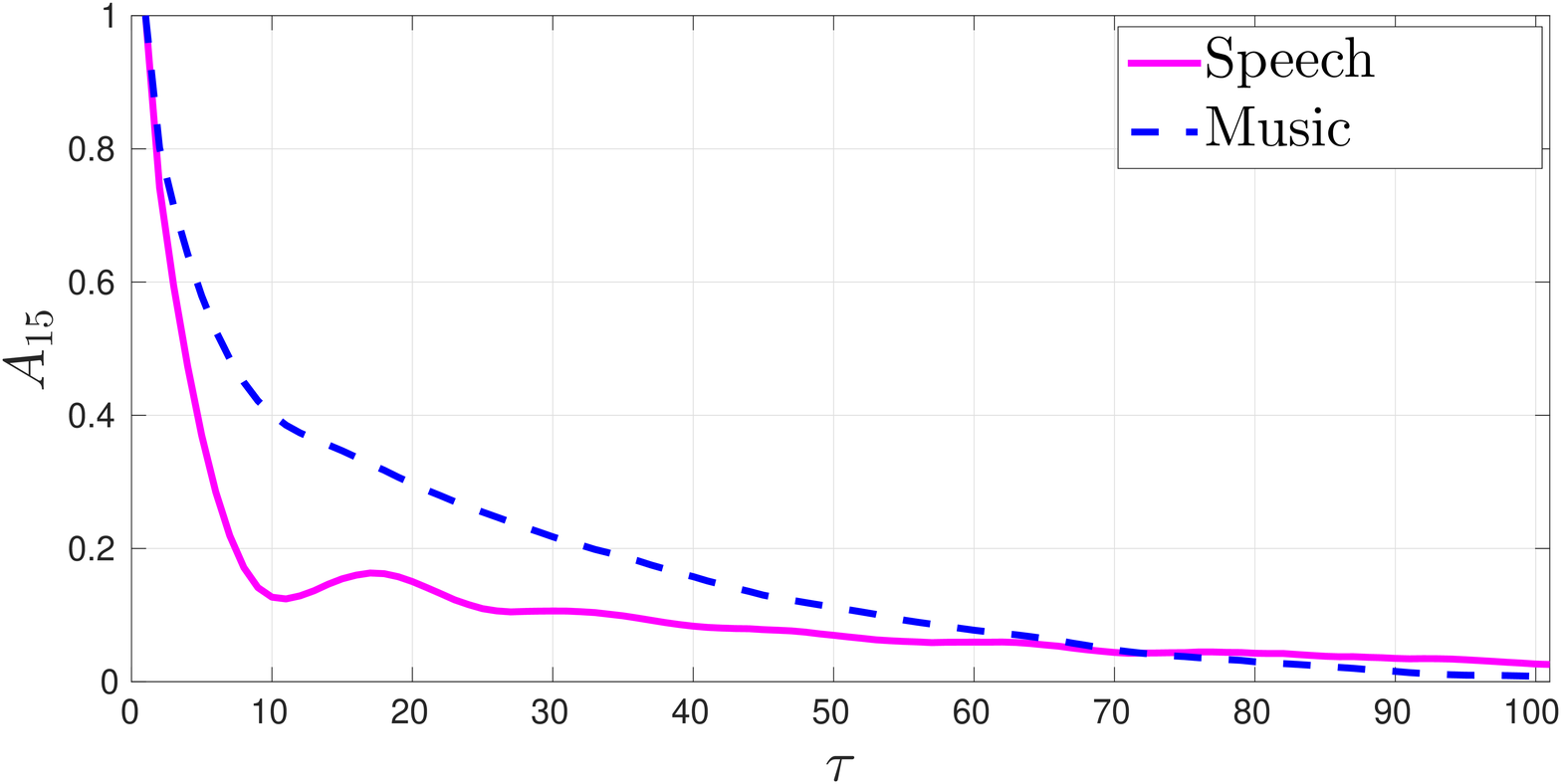}
\includegraphics[width=0.2\textwidth,keepaspectratio]{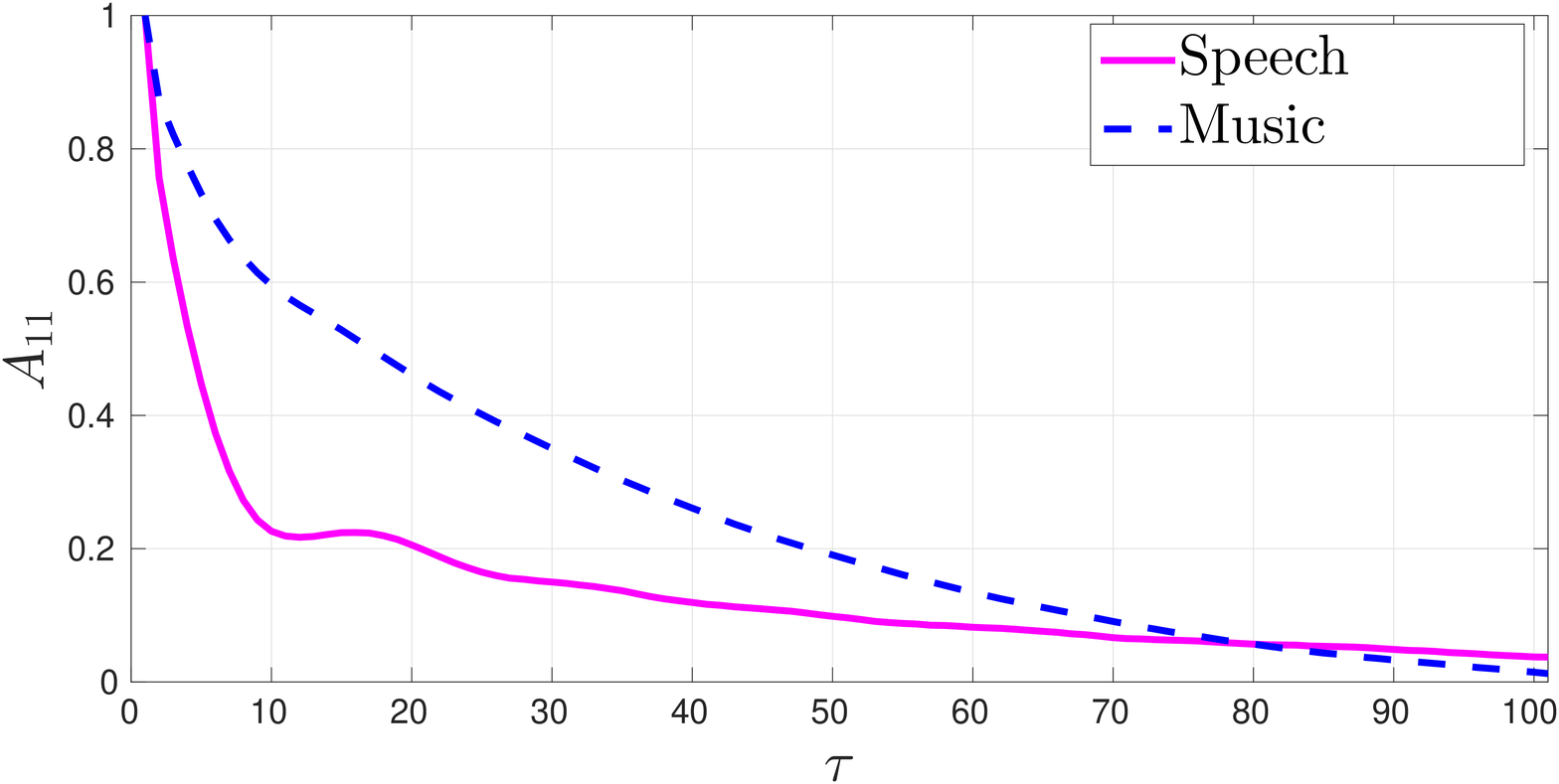}
\includegraphics[width=0.2\textwidth,keepaspectratio]{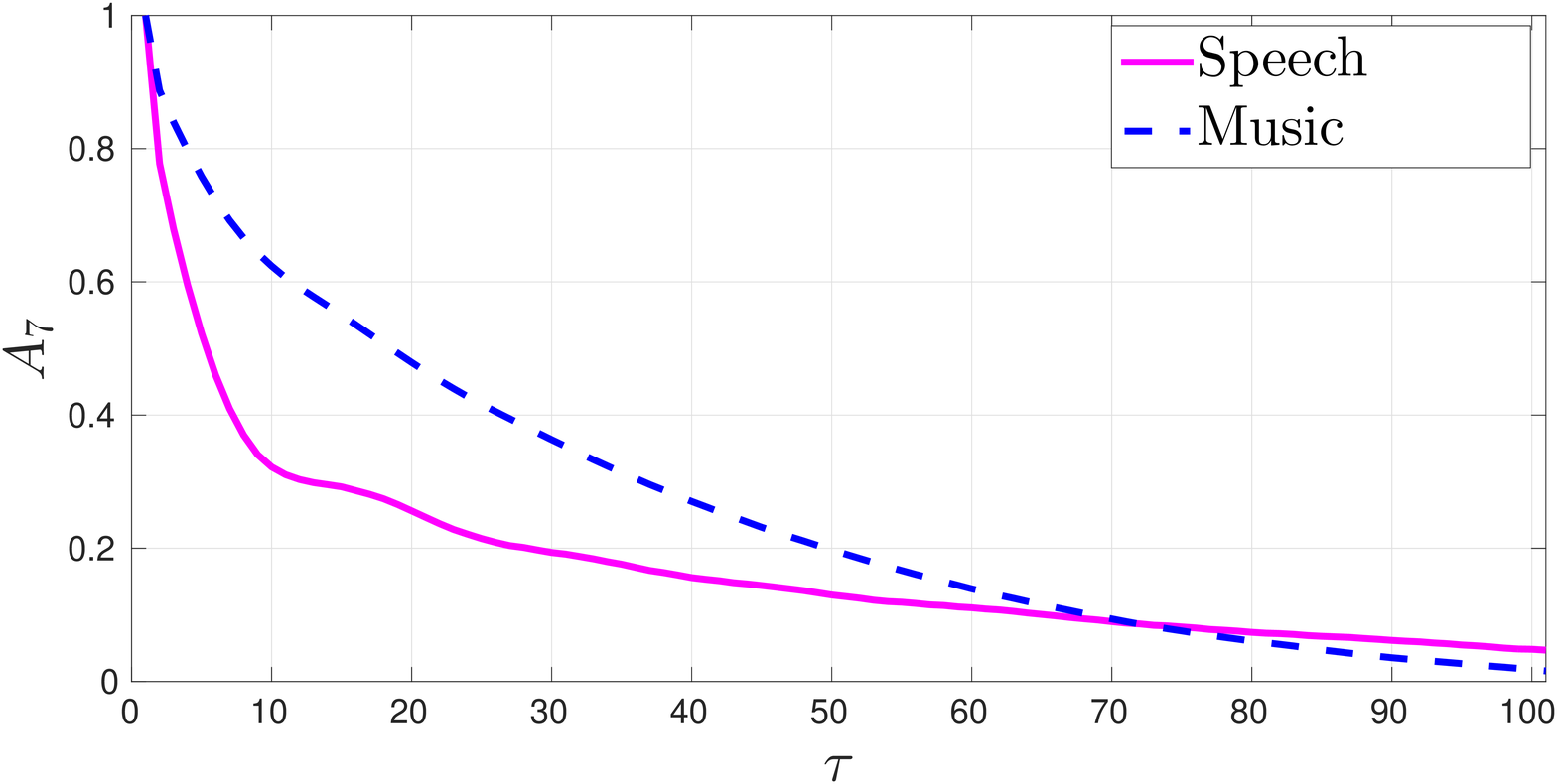}
\includegraphics[width=0.2\textwidth,keepaspectratio]{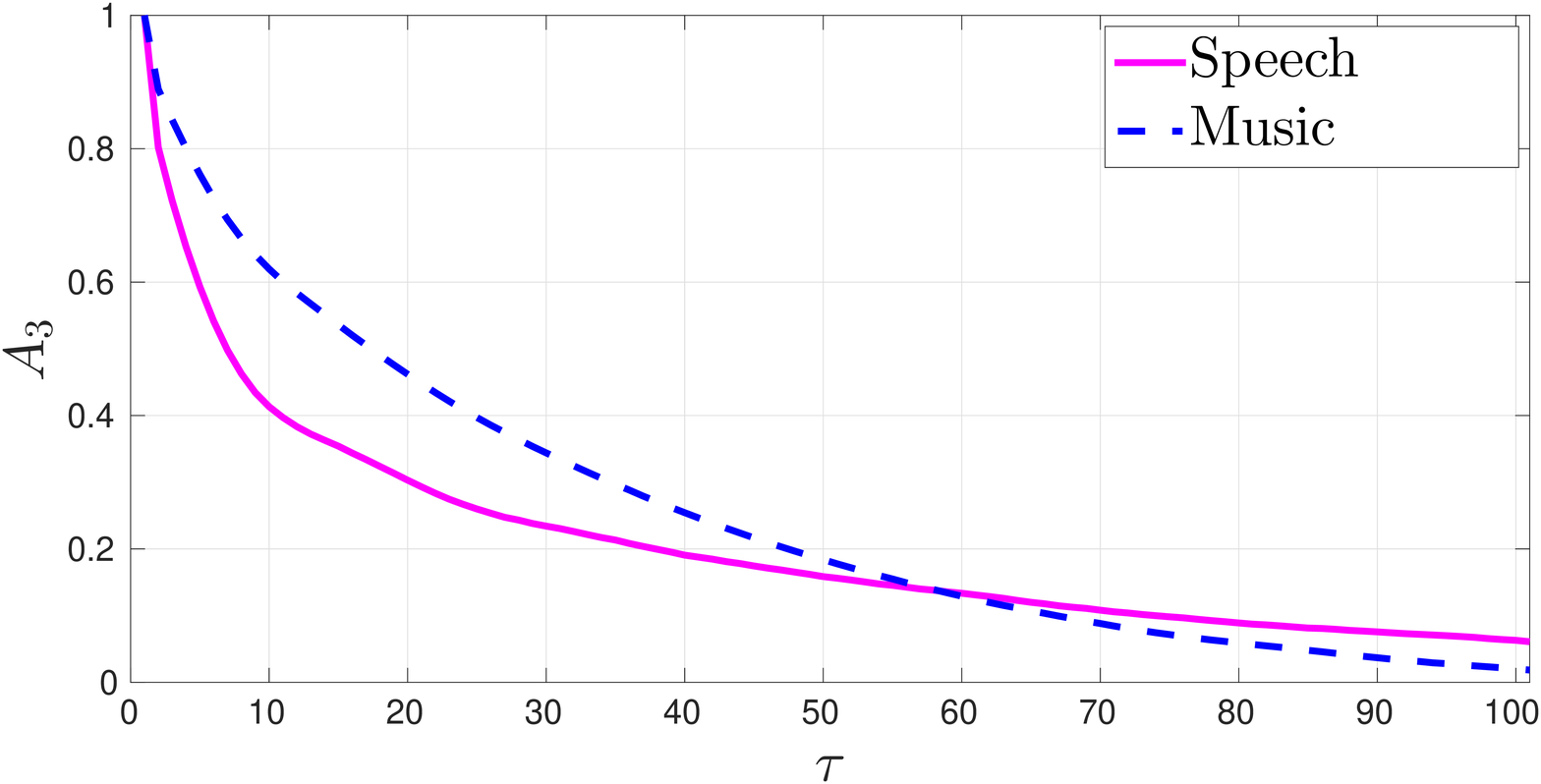}
\vspace{-0.5\baselineskip}
}
\centerline{
(a) 
\hspace{0.17\textwidth}	(b)
\hspace{0.17\textwidth}	(c)
\hspace{0.17\textwidth}	(d)
\hspace{0.17\textwidth}	(e)
}
\centerline{
\includegraphics[width=0.2\textwidth,keepaspectratio]{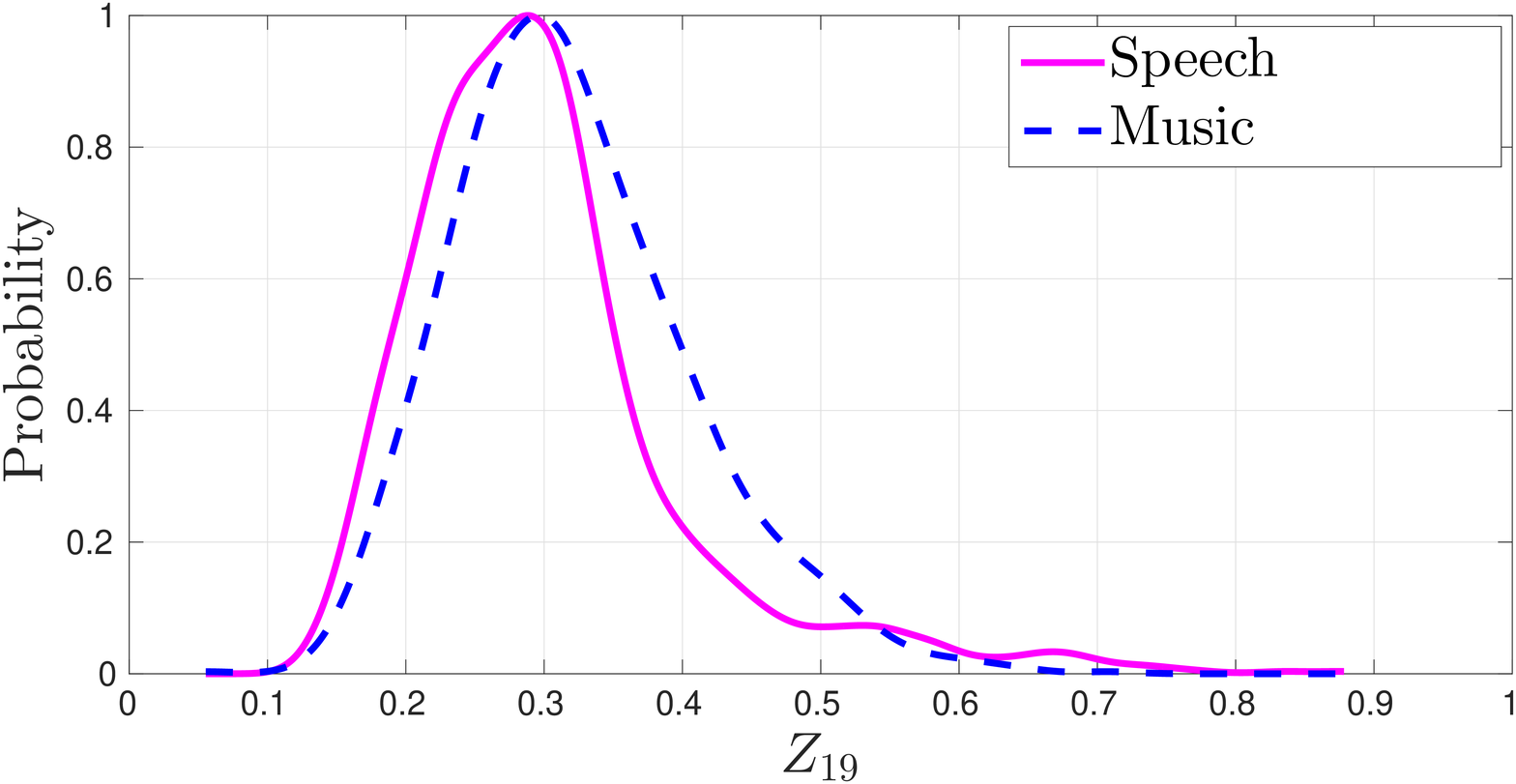}
\includegraphics[width=0.2\textwidth,keepaspectratio]{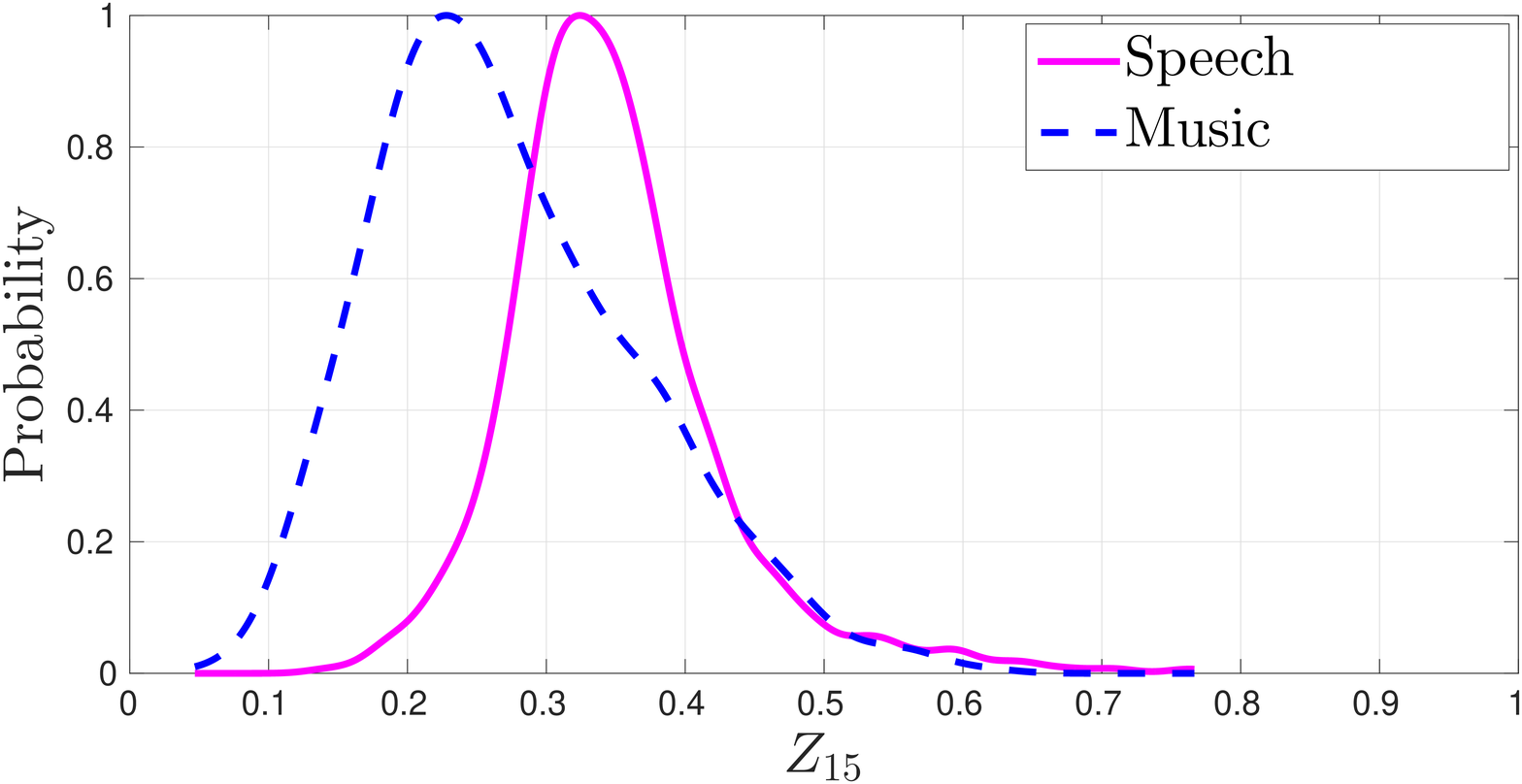}
\includegraphics[width=0.2\textwidth,keepaspectratio]{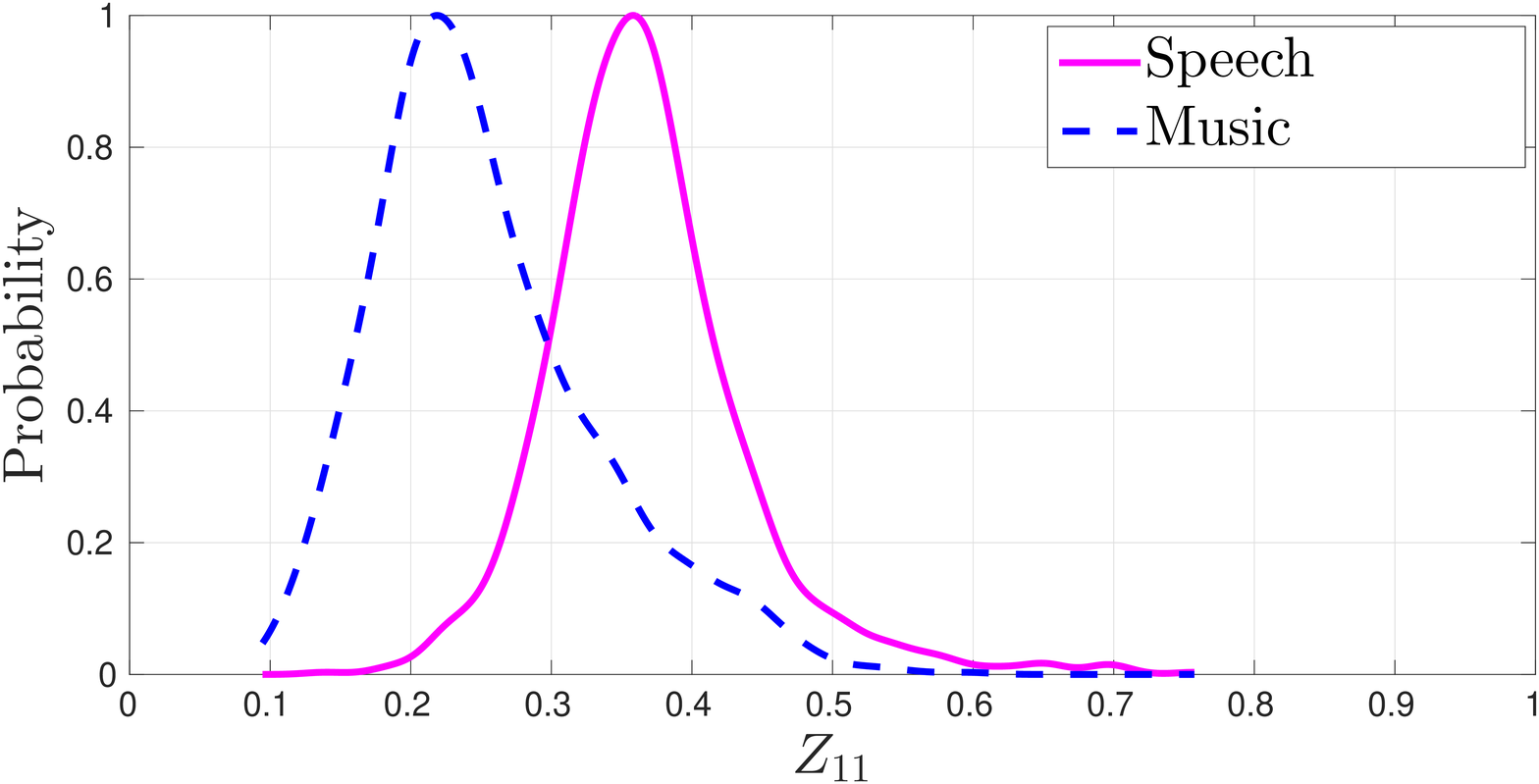}
\includegraphics[width=0.2\textwidth,keepaspectratio]{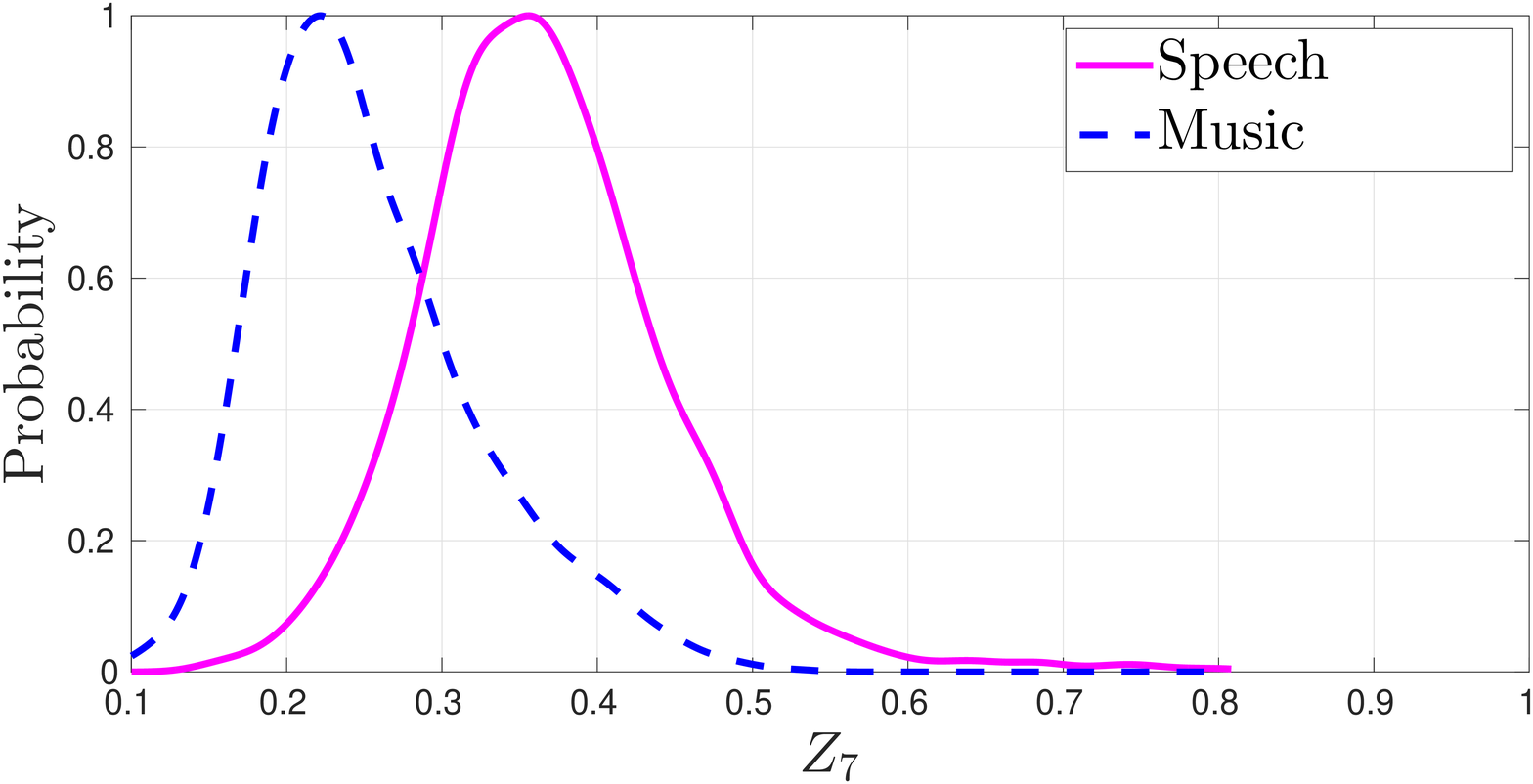}
\includegraphics[width=0.2\textwidth,keepaspectratio]{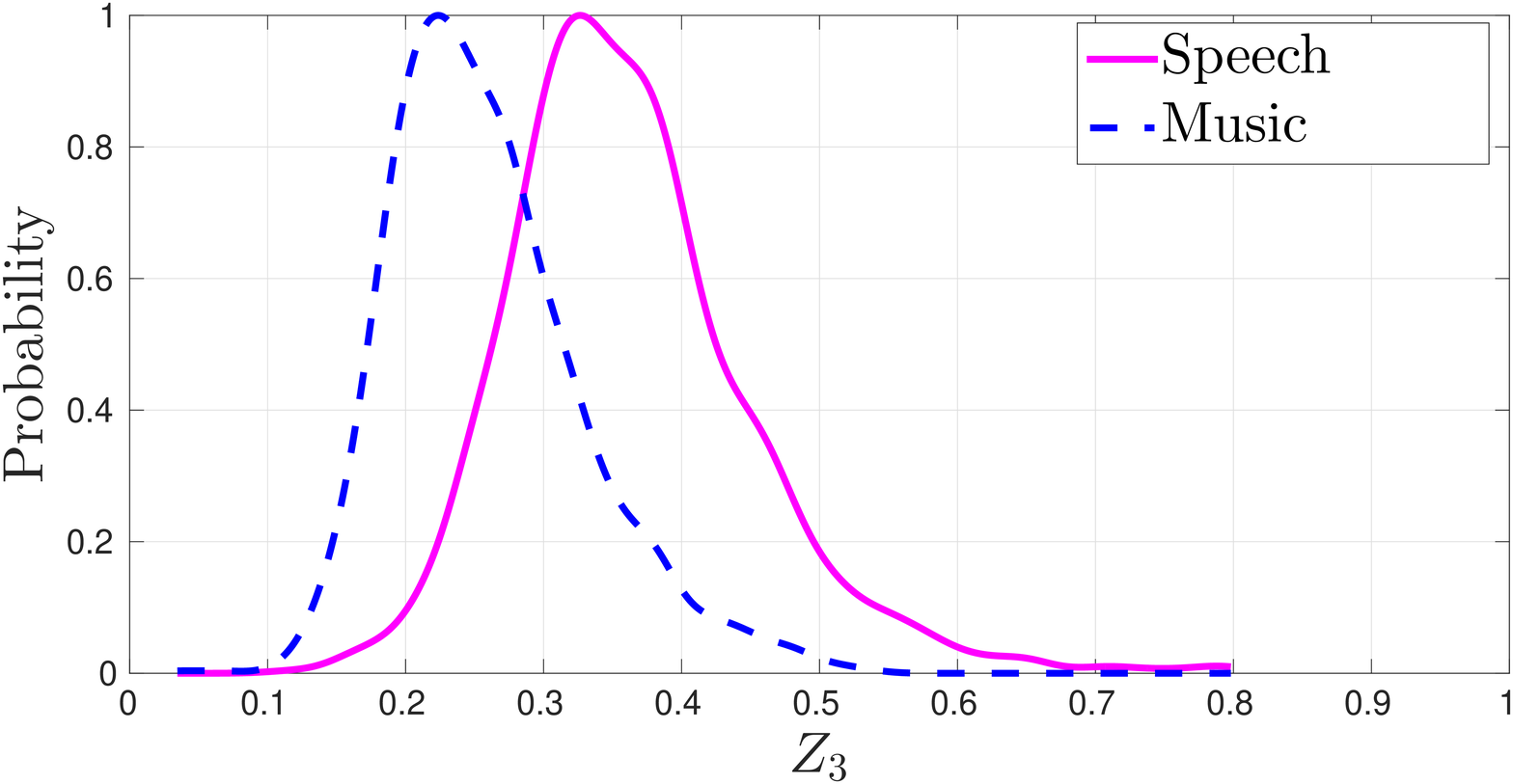}
\vspace{-0.5\baselineskip}
}
\centerline{
(f) 
\hspace{0.17\textwidth}	(g) 
\hspace{0.17\textwidth}	(h) 
\hspace{0.17\textwidth}	(i) 
\hspace{0.17\textwidth}	(j) 
}
\centerline{
\includegraphics[width=0.33\textwidth,keepaspectratio]{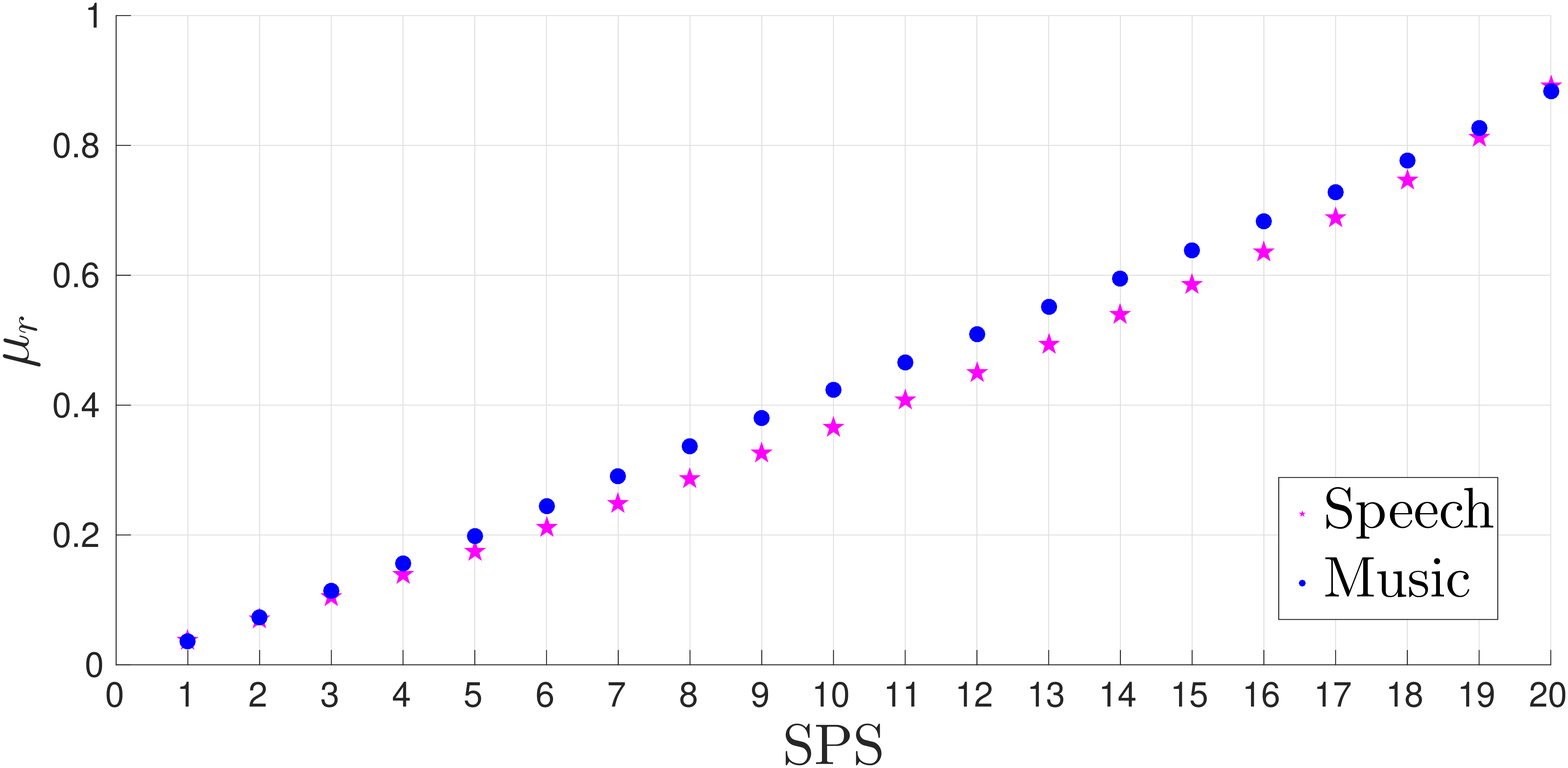}
\includegraphics[width=0.33\textwidth,keepaspectratio]{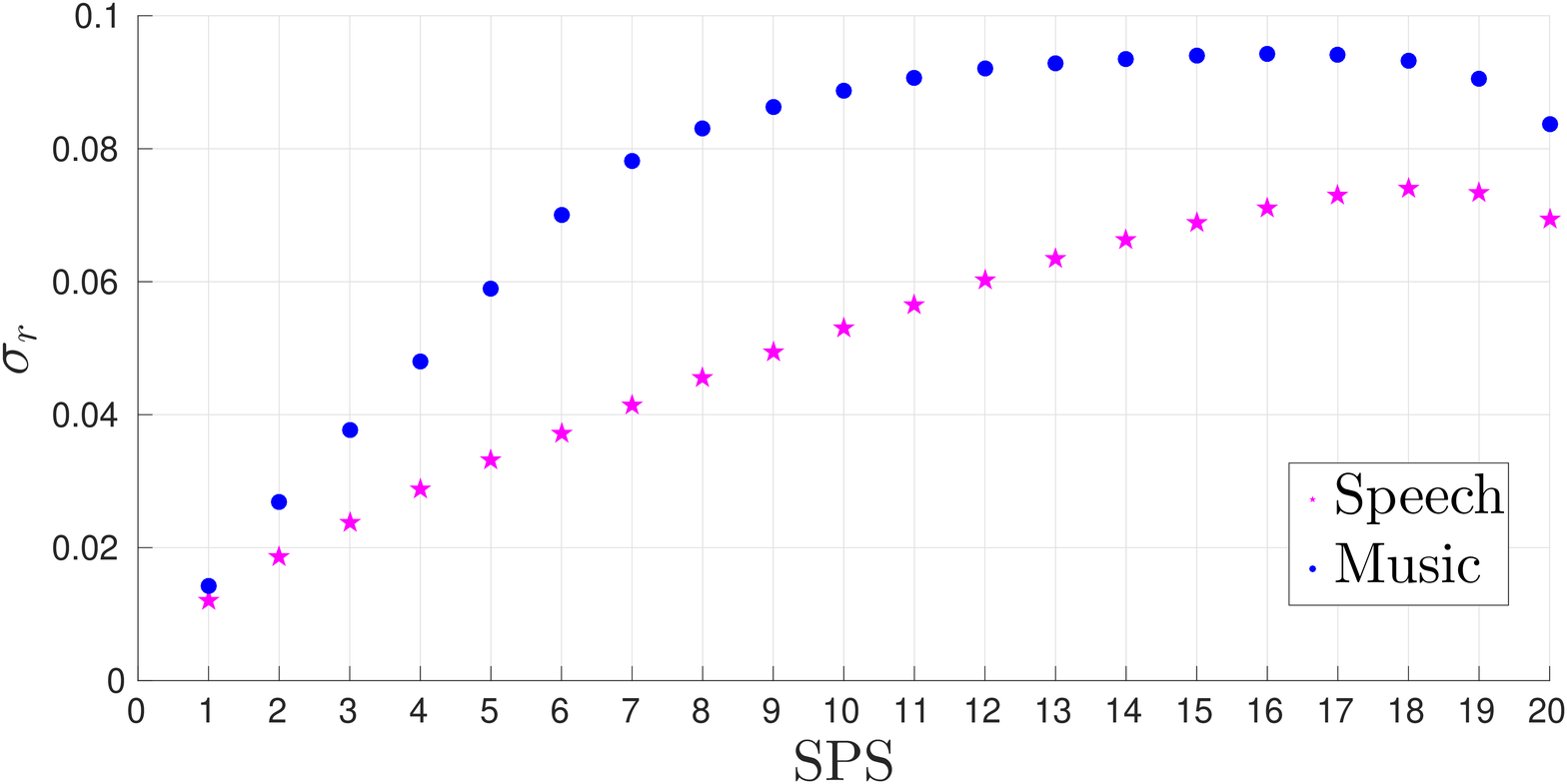}
\includegraphics[width=0.33\textwidth,keepaspectratio]{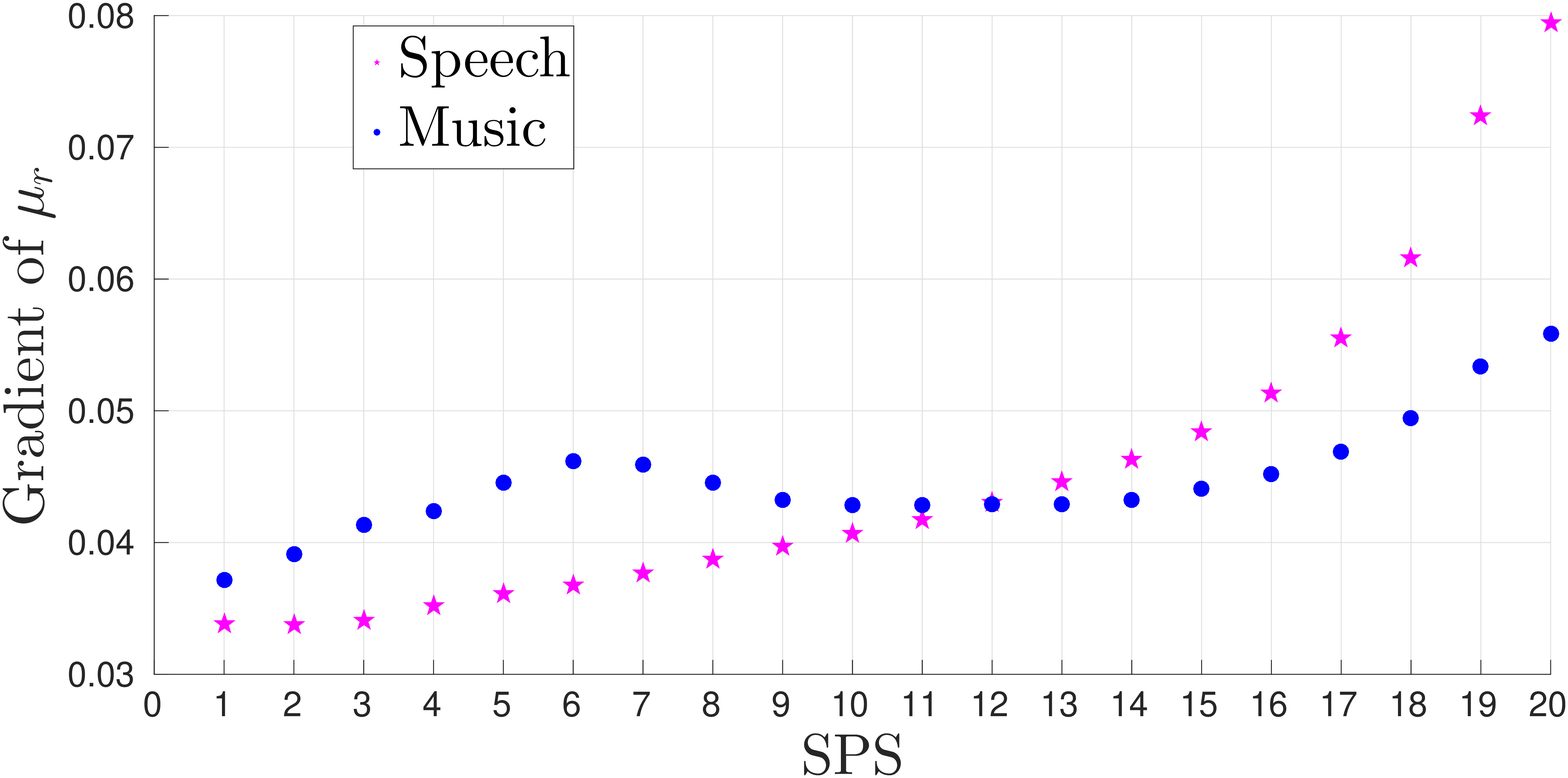}
\vspace{-0.5\baselineskip}
}
\centerline{
(k) 
\hspace{0.3\textwidth} (l)
\hspace{0.3\textwidth} (m)
}
\setlength{\belowcaptionskip}{-\baselineskip}
\caption{ Proposed features computed from the GTZAN dataset.
\textbf{(a)-(e)} show the trend of autocorrelation sequence $A_{r}$. Speech $A_{r}$ indicate presence of periodicity; 
\textbf{(f)-(j)} show the SPS-ZCR distribution. Speech in general have higher SPS-ZCR values than music; 
\textbf{(k)-(m)} show the values of $\mu_{r}$, $\sigma_{r}$ and $\Delta\mu_{r}$. Speech and music show distinct trends; 
Figures represent averaged behavior over the GTZAN data-set. SPS-ZCR and $A_{r}$ are shown only for $3^{rd}$, $7^{th}$, $11^{th}$, $15^{th}$ and $19^{th}$ SPS of speech and music.
}
\label{fig:SPS-features}
\end{figure*}

The audio segment $\mathbf{x}$ $\left( \mathbf{x}[n] \in \mathcal{R}; n=0, \ldots N_{s}-1 \right)$ is divided into $L$ overlapping frames $\mathbf{x}_{l}$ $\left(l=0,\ldots L-1 \right)$ of size $2N_{f}$. Let, $\mathbf{X}_{l}[k] = \sum\limits_{m = 0}^{2N_{f}-1} \mathbf{x}_{l}[m] e^{-jk\dfrac{2\pi}{2N_{f}}m}$ ($k = 0 \ldots 2N_{f}-1$) be the DFT of $\mathbf{x}_{l}$. These frames ($\mathbf{x}_{l}$) are sequences of real numbers. Hence, we consider only the first half of DFT coefficients (i.e. $\mathbf{X}_{l}[k]; k=0, \ldots N_{f} - 1$) from each frame. The proposed features are extracted in two stages and are described next. 

The first stage identifies the important spectral peaks present in each frame of the audio interval. The frequency locations of all spectral peaks in the $l^{th}$ frame are stored in a set $\mathbf{H}_{l}$. This set is constructed as

\vspace*{-\baselineskip}
\begin{equation}
\mathbf{H}_{l} = \left\lbrace k : 
\left[ \mathbf{X}_{l}[k-1] < \mathbf{X}_{l}[k] \right ] \wedge \\
\left[ \mathbf{X}_{l}[k] > \mathbf{X}_{l}[k+1] \right ]
\right\rbrace
\label{eq:PeakDetection}
\end{equation}

\noindent where $0 \leq k < (N_{f} - 1 )$. The number of spectral peaks ($|\mathbf{H}_{l}|$) varies in each frame varies. Thus, we retain at most $p$ prominent spectral peaks from each frame to construct the truncated set 

\begin{equation}
\nonumber
\mathbf{tH}_{l} = \left\{ k_{0}^{(l)}, k_{1}^{(l)}, \ldots k_{p-1}^{(l)}: \mathbf{X}_{l}[k_{0}] \geq \mathbf{X}_{l}[k_{1}] \geq \ldots \geq \mathbf{X}_{l}[k_{p}] \right\}
\end{equation}

\noindent However, if $|\mathbf{H}_{l}| = q < p$ then, the last frequency location $k_{q-1}$ is repeated $p-q$ times to maintain uniformity in cardinality of $\mathbf{tH}$ for all frames. The elements of $\mathbf{tH}_{l}$ are further sorted in descending order to construct the vector $\mathbf{pH}_{l} = \left[ k_{(0)}^{(l)} , k_{(1)}^{(l)} , \ldots k_{(p-1)}^{(l)} \right]$ ($k_{(0)}^{(l)} \geq k_{(1)}^{(l)} \geq \ldots \geq k_{(p-1)}^{(l)}$). These vectors ($\mathbf{pH}$) are used to construct a $p \times L$ peak sequence matrix $\mathbf{S}_{peak} = \left[ \mathbf{pH}_{0}^{T}, \ldots \mathbf{pH}_{L-1}^{T} \right]$ for an audio interval. Each row of $\mathbf{S}_{peak}$ is defined as a \emph{Spectral Peak Sequence} (SPS, henceforth). It is noteworthy that, the first row of $\mathbf{S}_{peak}$ corresponds to the SPS with highest frequency locations and the last row corresponds to one with lowest frequency locations.

In second stage, the proposed features are extracted from $\mathbf{S}_{peak}$. For notational convenience, the index $r$ ($0 \leq r < p$) will be used for referring to the $r^{th}$ row of $\mathbf{S}_{peak}$ or the $r^{th}$ SPS. Attributes derived from the $r^{th}$ SPS will also be indexed by $r$. This work proposes three different features derived from the SPS. These are \textbf{(a)} SPS Periodicity (\textbf{SPS-P}, henceforth), \textbf{(b)} SPS Zero Crossing Rate (\textbf{SPS-ZCR}, henceforth), and \textbf{(c)} SPS Standard Deviation, Centroid and its Gradient (\textbf{SPS-SCG}, henceforth). The following are computed from the SPS for feature extraction. Let $\mu_{r} = \dfrac{1}{L} \displaystyle \sum_{l=0}^{L-1} \mathbf{S}_{peak}[r][l]$ be the centroid frequency location of the $r^{th}$ SPS. These centroid frequencies are used to construct the zero-centered SPS $C_{r}$ such that $C_{r}[l] = \mathbf{S}_{peak}[r][l] - \mu_{r}$ ($l=0,\ldots L-1$). The auto-correlation sequence of $C_{r}$ can be estimated as $A_{r}[\tau] = \dfrac{1}{L} \displaystyle \sum\limits_{l=0}^{L-1-\tau}C_{r}[l]C_{r}[l+\tau]$ where, $\tau = 0, \ldots \mathcal{L}$ ($\mathcal{L} = \dfrac{L}{2}$ if $L$ is even and $\dfrac{L+1}{2}$ otherwise). One or more of these attributes are used to compute the proposed features.

\begin{description}[style=unboxed,leftmargin=0cm]

\item[SPS-Periodicity] -- It is well known that quasi-periodic voiced sounds constitute a major part of the speech signals \cite{Biswas_IJST2014, Kawahara_ICASSP2013}. Whereas, music is created by musicians with their personalized styles of arranging sound items from multiple instruments. Hence, music signals need not necessarily have a periodic nature. Figures~\ref{fig:SPS-features}(a)-(e) show the average trends in autocorrelation sequences of different speech and music SPS estimated from the GTZAN dataset. Presence of peaks (other than the first one) in autocorrelation sequence of a signal indicates its periodicity. Such peaks are observed in autocorrelation sequences of SPS of speech but, not in that of music. This motivated us to exploit the periodicity of SPS as feature for speech-music discrimination. Periodicity of the $r^{th}$ SPS is estimated using its auto-correlation sequence. The peak locations $\tau^{(r)}$ of $A_{r}$ are detected (Equation~\ref{eq:PeakDetection}) and stored in a set $\mathbf{T}_{r} = \left\{ \tau_{0}^{(r)}, \tau_{1}^{(r)}, \ldots \right\}$ ($|\mathbf{T}_{r}| < \mathcal{L}$) in an ascending order. We compute the quantities $\Delta\tau_{u}^{(r)} = \tau_{u}^{(r)} - \tau_{u-1}^{(r)}$ ($u = 1, \ldots |\mathbf{T}_{r}| - 1$). The variance $V_{r}$ of these quantities $\{\Delta\tau_{u}^{(r)}\}$ provides an estimate of the periodicity of the $r^{th}$ SPS. The feature SPS-P is constructed as a $p$ dimensional vector such that $SPS\textit{-}P = [V_{0}, \ldots V_{p-1}]$.

\item[SPS-Zero Crossing Rate] -- Audio signals are non-stationary. Thus, spectral peaks in a certain SPS may correspond to different frequency locations within the spectra of audio frames in an interval. Hence, without any loss of generality, we can assume that spectral peak sequences contain varying values. The Zero Crossing Rate (ZCR) provides a gross estimate of average frequency of time-series data \cite{Shete_JSVP2014}. We propose to compute the ZCR of each SPS to estimate their average frequency and use this as a feature for CSM. The ZCR $\left(Z_{r}\right)$ of the $r^{th}$ zero-centered SPS is computed as $Z_{r} = \dfrac{1}{2L} \displaystyle \sum\limits_{l=0}^{L-1} \abs{sgn\left(C_{r}[l]\right) - sgn\left(C_{r}[l-1]\right) }$ where, $sgn(\bullet)$ is the signum function. SPS-ZCR feature is constructed as a $p$ dimensional vector such that $SPS\textit{-}ZCR = [Z_{0}, \ldots Z_{p-1}]$. Figures~\ref{fig:SPS-features}(f)-(j) show the distributions of ZCR values for different SPS of speech and music. We observe that, ZCR of lower-frequency SPS (e.g. $Z_{19}$, Figure~\ref{fig:SPS-features}(f)) exhibit significant overlap between the ZCR distribution of two classes. However, this overlap reduces as music SPS-ZCR values gradually decrease (compared to that of speech) for higher-frequency spectral peak sequences ($Z_{15}$ to $Z_{3}$, Figures~\ref{fig:SPS-features}(g)-(j)).
In general, speech SPS-ZCR values are higher than that of music, indicating that speech SPS values vary more than that of music. Hence, this property can be exploited as a discriminator between the two classes.

\item[SPS-Standard Deviation, Centroid and its Gradient] -- We believe that the frequency locations in any $r^{th}$ SPS are category specific (i.e. either speech or music). This motivated us to propose a set of features based on the statistical properties of the spectral peak sequences. These statistical attributes include the centroid $\mu_{r}$ and standard deviation $\sigma_{r}=\sqrt{ \dfrac{1}{L} \sum\limits_{l=0}^{L-1} \left(\mathbf{S}_{peak}[r][l] - \mu_{r} \right)^2 }$ of the $r^{th}$ SPS. Also, the rates of change of $\mu_{r}$ (with respect to $r$) exhibit distinct trends for both speech and music. We compute the gradient $\Delta\mu_{r} = \dfrac{1}{2} \left(\mu_{r+1} - \mu_{r-1}\right)$ for representing this trend. Thus, we propose the SPS-SCG feature as a $3p$ dimensional vector given by $SPS\textit{-}SCG = [\mu_{0}, \ldots \mu_{p-1}, \sigma_{0}, \ldots \sigma_{p-1}, \Delta\mu_{0}, \ldots \Delta\mu_{p-1}]$. Here, $\Delta\mu_{0} = \left(\mu_{1} - \mu_{0}\right)$ and $\Delta\mu_{p-1} = \left(\mu_{p-1} - \mu_{p-2}\right)$. Figure~\ref{fig:SPS-features}(k)-(m) show the trends of SPS-SCG features averaged over several audio intervals for both speech and music (GTZAN dataset).

\end{description}

The proposed features capture prominent spectral information in the first stage and temporal variations are characterized in the second stage. Binary classifiers are learned on these proposed features. In this proposal, we have experimented with Gaussian mixture models (GMM), support vector machines (SVM) and random forest (RF) classifiers. The results of our experiments with these tempo-spectral features are presented next.

\begin{table}[t]
\caption{Performance of baseline approaches and individual features on GTZAN dataset. Additionally, performances of early and late fusion of proposed features are also presented. Experiments are performed with GMM, SVM and Random Forest. The classifier parameters are optimized by grid-search. SPS-SCG with SVM has better performance compared to baseline approaches and other features.}
\label{table:PrpsdFeatPerf}
\renewcommand{\arraystretch}{1.3}
\resizebox{\columnwidth}{!}{
\centering
\begin{tabular}{
C{0.25\columnwidth}
C{0.25\columnwidth}
C{0.25\columnwidth}
C{0.25\columnwidth}
}

\cline{2-4}
& \textbf{GMM} & \textbf{Random Forest} & \textbf{SVM} \\
\hline

\textbf{Khonglah-FS}	&	$0.91 \left(0.02\right)$	&	$0.93 \left(0.02\right)$	&	$0.93 \left(0.01\right)$ \\
\hline

\textbf{Sell-FS}	&	$0.94 \left(0.01\right)$	&	$0.95 \left(0.01\right)$	&	$0.95\left(0.01\right)$ \\
\hline

\textbf{MFCC}	&	$\mathbf{0.95 \left(0.01\right)}$	&	$0.92 \left(0.02\right)$	&	$0.97 \left(0.01\right)$ \\
\hline

\textbf{SPS-P}	&	$0.83 \left(0.05\right)$	&	$0.86 \left(0.04\right)$	&	$0.84 \left(0.05\right)$ \\
\hline

\textbf{SPS-ZCR}	&	$0.81 \left(0.04\right)$	&	$0.84 \left(0.04\right)$	&	$0.87 \left(0.03\right)$ \\
\hline

\textbf{SPS-SCG}	&	$0.93 \left(0.01\right)$	&	$\mathbf{0.95 \left(0.01\right)}$	&	$\mathbf{0.98 \left(0.00\right)}$ \\
\hline

\textbf{SPS-EF}	&	$0.93 \left(0.02\right)$	&	$0.95 \left(0.01\right)$	&	$0.98 \left(0.00\right)$ \\
\hline

\textbf{SPS-LF}	&	$0.91 \left(0.02\right)$	&	$0.95 \left(0.01\right)$	&	$0.92 \left(0.02\right)$ \\
\hline

\end{tabular}
}
\vspace{-\baselineskip}
\end{table}

\begin{figure}[t]
\centerline{\includegraphics[width=0.99\columnwidth,keepaspectratio]{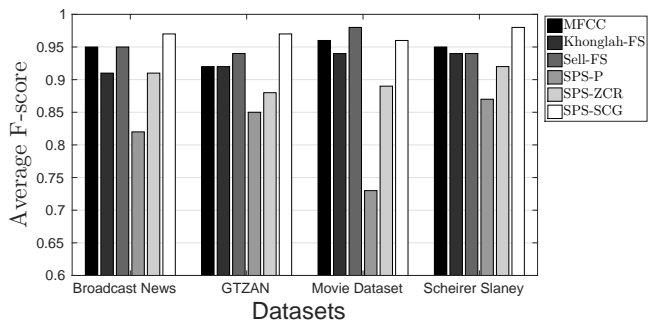}}
\setlength{\belowcaptionskip}{-\baselineskip}
\caption{The performance of baseline and proposed features on four data-sets using SVM (with radial basis function kernel) classifier. Among proposed features, SPS-SCG has best performance on three out of four datasets.}
\label{fig:PerformanceVariance}
\vspace{-0.5\baselineskip}
\end{figure}

\vspace*{-0.5\baselineskip}
\section{Experiments and Results}
\label{sec:Experiments}

The proposed approach is validated on four datasets. These are \textbf{(a)} GTZAN Music/Speech collection \cite{Tzanetakis_TSAP2002}, \textbf{(b)} Scheirer-Slaney Music-Speech Corpus \cite{Scheirer_ICASSP1997}, \textbf{(c)} Movie dataset, \textbf{(d)} TV News Broadcast dataset. The later two datasets are created by us and are available on request for non-commercial usage. The movie dataset consists of $5s$ clips of pure speech and pure music from old Bollywood movies. The TV News Broadcast dataset contains $5s$ clips of speech and non-vocal music recorded from Indian English news channels. 

Our proposal is benchmarked against the following three baseline approaches. First, the method proposed by Khonglah et al. in \cite{Khonglah_DSP2016} (Khonglah-FS). The authors propose that speech specific features like Normalized Autocorrelation Peak Strength of the Zero Frequency Filtered Signal, the Peak-to-Sidelobe Ratio from Hilbert Envelope of the LP residual, Log-Mel Spectrum Energy, and 4-Hz Modulation Energy etc. are better in characterizing speech and hence, good discriminators from music. The second approach proposed by Sell et al. \cite{Sell_ICASSP2014} (Sell-FS) uses novel chroma based features that represent music tonality for better speech-music classification. Third, the $13$ MFCC coefficients \cite{Mezghani_AICCSA2016} (MFCC) are considered as features as these are widely used in most speech processing applications.

For all our experiments, we have chosen audio intervals of $1~s$ duration. From each audio interval, we have drawn frames of $30~ms$ duration with a shift of $1~ms$. Features are extracted from each audio interval. Accordingly, each audio interval is classified as either speech or music. The number of prominent peaks $p$ is empirically selected and is set to $p = 20$ for all our experiments. We have used MATLAB toolboxes for realizing the GMM and RF based classifiers. The lib-SVM toolbox \cite{Chang_ATIST2011} is used for SVM with radial basis function kernel based classifier. The classifier parameters are optimized by grid-search. The training and test data are chosen in a ratio of $70:30$. The experiments are repeated $20$ times. The mean and variances of F-scores of these independent trials are reported.

The performance of baseline approaches and individual features from our proposal (on GTZAN only) are presented in Table~\ref{table:PrpsdFeatPerf}. SPS-P and SPS-ZCR fail to outperform the baseline approaches. However, SPS-SCG provides a significant improvement over the best baseline. Additionally, we have experimented with early and late feature fusion schemes for our proposal. However, no significant improvement was observed over the performance of SPS-SCG. The comparative performance analysis of proposed features and baseline approaches (with SVM only) for all four datasets are shown in Figure~\ref{fig:PerformanceVariance}. The SPS-SCG features with SVM classifier provides the best performance for GTZAN, Scheirer Slaney and TV News Broadcast dataset. However, it has second best performance for the Movie data-set. Thus, the experimental results establish that the proposed features can effectively capture the time-frequency characteristics of speech and music while discriminating one from another.

\vspace{-0.5\baselineskip}
\section{Conclusion}
\label{sec:Conclusion}

This work proposes a novel two-stage feature extraction scheme for representing the time-frequency characteristics of an audio interval. In the first stage, we detect the frequency locations of $p$ prominent spectral peaks for each frame in an audio interval. These peak locations are stored as columns in a matrix $\mathbf{S}_{peak}$. The rows of this matrix are defined as the $p$ spectral peak sequences (SPS) that characterize the audio interval. The proposed features are computed in the second stage by treating each SPS as temporal sequence. We estimate the periodicity (SPS-P), ZCR (SPS-ZCR), standard deviation, centroid and its gradient (collectively, SPS-SCG) as features of each SPS. The performance of our proposal is benchmarked on four datasets and against three baseline approaches. The proposed features are deployed with GMM, SVM and Random Forest based classifiers. Among the proposed features, SPS-SCG (with SVM) has better performance compared to baseline approaches and other features on three datasets.

The spectral peak sequences are prominent peak locations (integer values) of frame spectra. This feature can be extended to incorporate sequences of other attributes of frame spectra. The present work focuses on ZCR, periodicity and a few statistical attributes of the spectral peak sequences. This can be further enhanced by considering other temporal sequence features. The proposed features are applied to the domain of speech-music classification. This work can be extended to deploy an enhanced set of these features for effective discrimination of speech, music and multiple categories of environmental sounds.

\bibliographystyle{IEEEtran}
\bibliography{references}

\end{document}